\documentclass[12pt]{article}
\usepackage{graphicx}

\newcommand\pubnumber{}
\newcommand\pubdate{\today}

\newcommand{\gevsq}{\ensuremath{\mathrm{\,Ge\kern -0.1em V^2\!}}}
\def\mainz{Institut f\"ur Physik (THEP), Johannes Gutenberg-Universit\"at\\
      D-55099 Mainz, Germany}
\def\glasgow{SUPA, School of Physics and Astronomy, University of Glasgow, Glasgow, G12~8QQ, UK}
\def\vienna{Institut f\"ur Hochenergiephysik, \"Osterreichische
      Akademie der Wissenschaften, A-1050 Wien, Austria}

\def\Title#1{\begin{center} {\Large #1 } \end{center}}
\def\Author#1{\begin{center}{ \sc #1} \end{center}}
\def\Address#1{\begin{center}{ \it #1} \end{center}}

\newcommand\pubblock{\rightline{\begin{tabular}{l} \pubnumber\\
         \pubdate  \end{tabular}}}
\newenvironment{Abstract}{\begin{quotation}  }{\end{quotation}}
\newenvironment{Presented}{\begin{quotation} \begin{center} 
             PRESENTED AT\end{center}\bigskip 
      \begin{center}\begin{large}}{\end{large}\end{center} \end{quotation}}
\def\Acknowledgements{\bigskip  \bigskip \begin{center} \begin{large}
             \bf ACKNOWLEDGEMENTS \end{large}\end{center}}
\def\beq{\begin{equation}}
\def\eeq#1{\label{#1}\end{equation}}
\def\eeqn{\end{equation}}

\def\beqa{\begin{eqnarray}}
\def\eeqa#1{\label{#1}\end{eqnarray}}
\def\eeqan{\end{eqnarray}}

\let\bar=\overbar

\def\Dslash{\not{\hbox{\kern-4pt $D$}}}
\def\dslash{\not{\hbox{\kern-2pt $\del$}}}

\def\msb{{\bar{\ssstyle M \kern -1pt S}}}

\begin{document}
\begin{titlepage}
\pubblock

\vfill
\Title{CKM2010 Working Group II Summary:\\
  Determination of $|V_{cs}|$, $|V_{cd}|$, $|V_{cb}|$ and $|V_{ub}|$}
\vfill
\Author{Jack Laiho}
\Address{\glasgow}
\Author{Ben~D.~Pecjak}
\Address{\mainz}
\Author{Christoph Schwanda}
\Address{\vienna}
\vfill
\begin{Abstract}
  We review the progress on the determination of the CKM matrix
  elements $|V_{cs}|$, $|V_{cd}|$, $|V_{cb}|$, $|V_{ub}|$ and heavy
  quark masses presented at the 6th International Workshop on the CKM
  Unitarity Triangle.
\end{Abstract}
\vfill
\begin{Presented}
Proceedings of CKM2010, the 6th International Workshop on the CKM
Unitarity Triangle, University of Warwick, UK, 6-10 September 2010
\end{Presented}
\vfill
\end{titlepage}
\setcounter{footnote}{0}

\section{Introduction}

The CKM matrix elements must be determined precisely in order to
constrain physics beyond the Standard Model.  This working group
report focuses on the most up-to-date results from theory and
experiment used to obtain $|V_{cs}|$, $|V_{cd}|$, $|V_{cb}|$, and
$|V_{ub}|$.  We mainly concentrate on results from semi-leptonic $b$
and $c$ decays, though we also discuss leptonic decays of $B$ and $D$
mesons and the determinations of $b$ and $c$ quark masses.

\section{\boldmath Semi-leptonic $D$~decays and determination of
  $|V_{cs}|$ and $|V_{cd}|$}

Semi-leptonic $D$ meson decays provide an opportunity to test lattice
QCD calculations of the form factors $f^{K,\pi}_+(q^2=0)$ if one
assumes Standard Model CKM unitarity.  One can also turn this test
around, using the lattice calculations of the form factors to directly
determine the CKM matrix elements $|V_{cs}|$ and $|V_{cd}|$, thus
testing the Standard Model via second row and second column unitarity.
There are currently three groups using different lattice formulations
to calculate properties of semi-leptonic $D$ decays.  The HPQCD
Collaboration has recently published an unquenched lattice result for
the $D\to K\ell\nu$ form factor $f^K_+(0)$ using Highly Improved
Staggered Quarks (HISQ) and used it to extract a value of $|V_{cs}|$
from experiment~\cite{Na:2010uf}.  The older unquenched lattice
calculation from the Fermilab Lattice and MILC Collaborations
\cite{fnallqcd} has larger errors, but includes results at non-zero
$q^2$, as well as the $D\to \pi\ell\nu$ form factor $f^\pi_+(q^2)$.  A
preliminary lattice calculation of $f^{K,\pi}_+(q^2)$ from the ETM
Collaboration with a quenched strange quark is also available and is
in good agreement with the other two results
\cite{DiVita:2011py}. Competitive results for the $D\to K$ and
$D\to\pi$~form factors are also obtained from light-cone sum
rules~\cite{Khodjamirian:2009ys}.

At CLEO-c, $D\bar D$~meson pairs are produced at threshold through the decays
$e^+e^-\to\psi(3770)\to D\bar D$ at a center-of-mass energy (c.m.) near
3.770~GeV. The integrated luminosity of the $\psi(3770)$~sample is
818~pb$^{-1}$ corresponding to about 5.4 million $D\bar D$~events. By
reconstructing the hadronic decay of one $D$ (tag side), the
4-momentum of the second charmed meson (signal side) is known. This
allows to reconstruct a semileptonic decay with no kinematic
ambiguity. For $D_s$~decays, CLEO-c uses a data sample taken at
$\sqrt{s}=4.170$~GeV equivalent to 600~pb$^{-1}$. The $D_s$~mesons
used are from the reactions $e^+e^-\to D^{*+}_s D^-_s$ or $D^+_s
D^{*-}_s$. At the $B$~factories Belle and BaBar, the
production of charmed mesons is accompanied by additional
particles from fragmentation. Belle performed a tagged analysis with
full reconstruction of events $e^+e^-\to
D^{(*)}_\mathrm{tag}D^{*-}_\mathrm{sig}X$. This method gives high
$q^2$~resolution at the price of low reconstruction efficiency. BaBar
has chosen a different approach and reconstructs the signal charm
meson only, the neutrino energy being evaluated from the rest of the
event. This method has much higher efficiency but the branching
fraction measurement has to be done using a normalization channel.

\begin{figure}[bptb]
\centering
  \includegraphics*[width=2.2in]{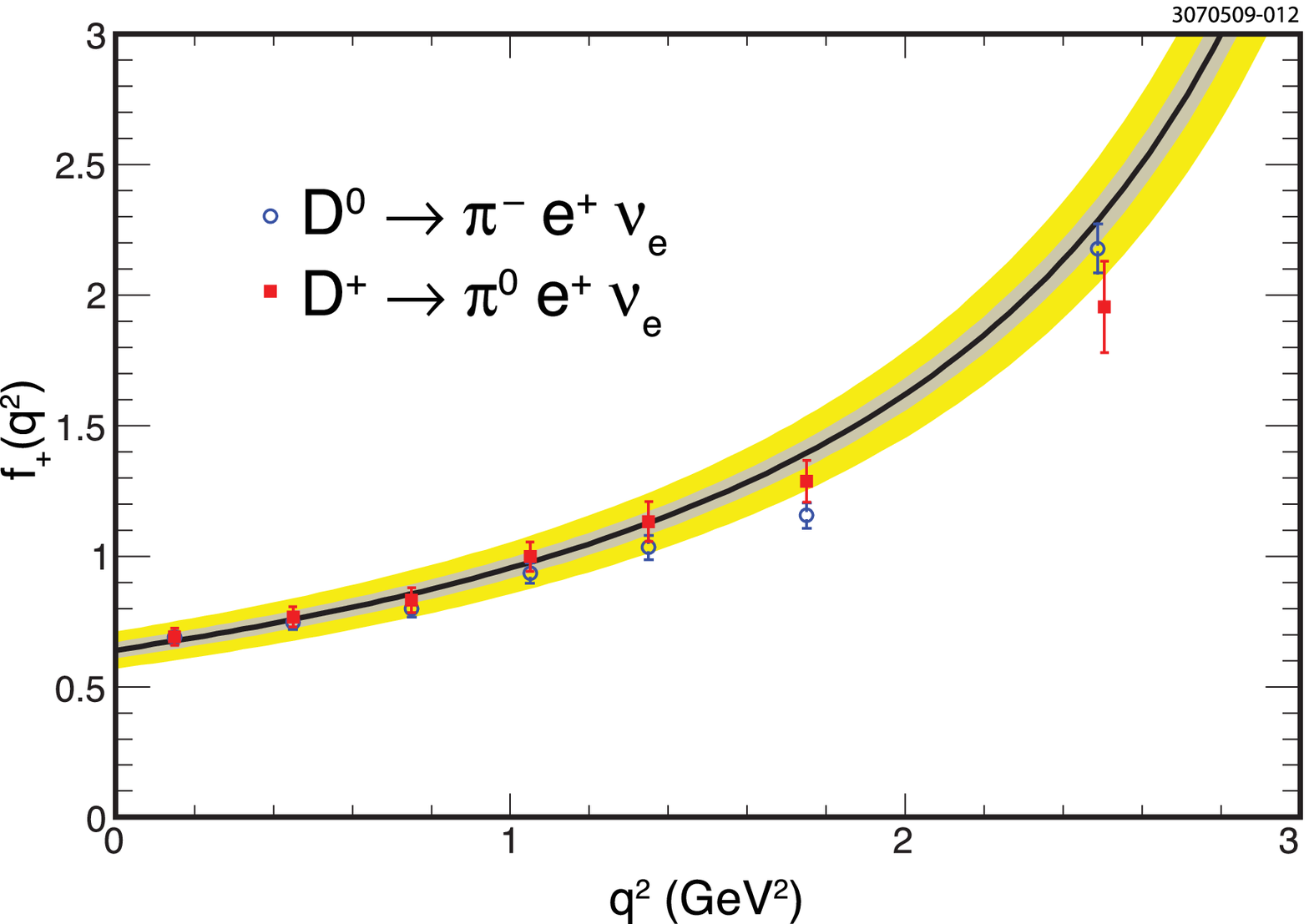}
  \includegraphics*[width=2.2in]{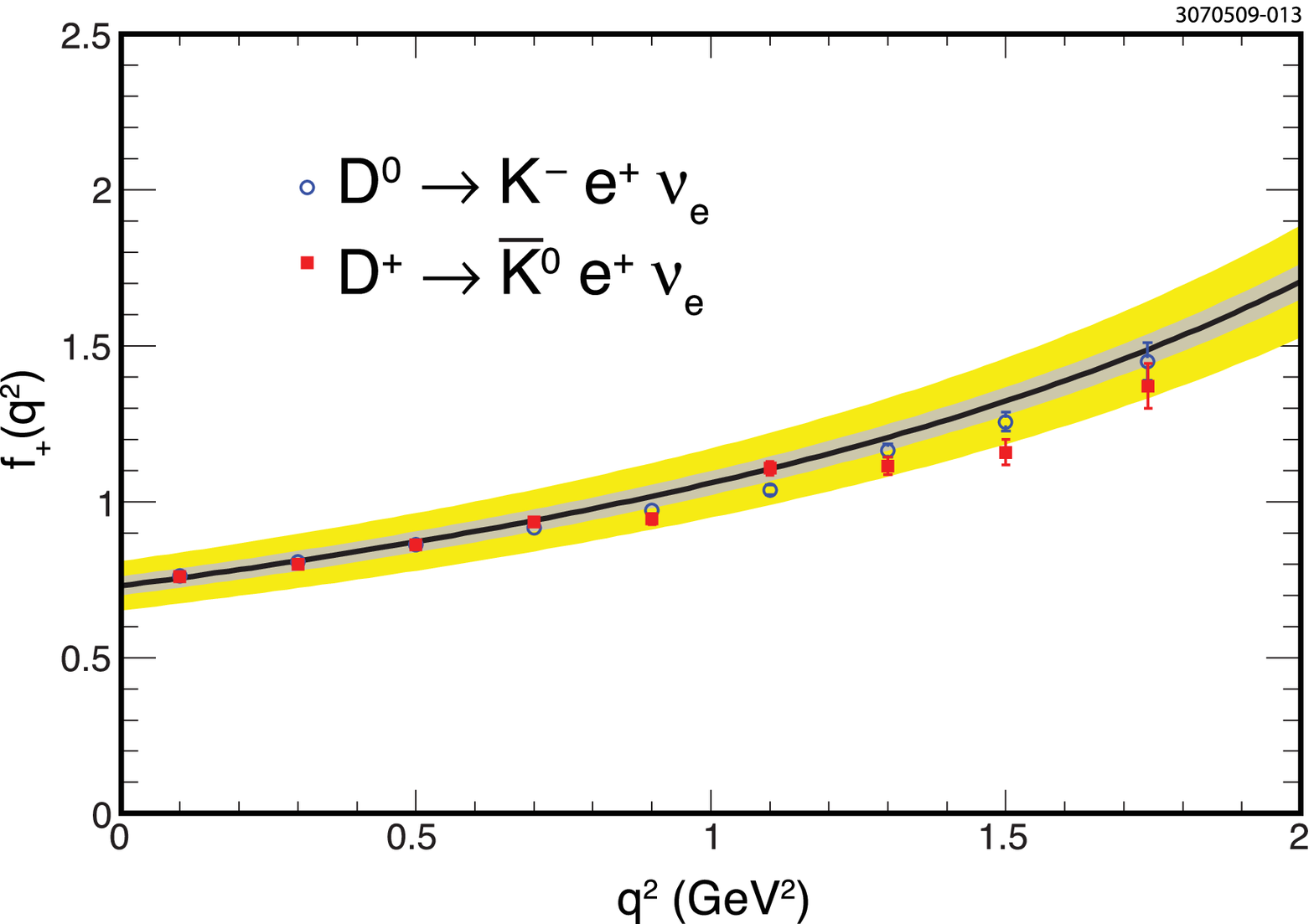}
  \caption{$f_+(q^2)$ comparison between isospin conjugate modes and
    with LQCD calculations~\cite{fnallqcd}. The solid lines represent
    LQCD fits to the modified pole model. The inner bands show LQCD
    statistical uncertainties, and the outer bands the sum in
    quadrature of LQCD statistical and systematic uncertainties.}
  \label{fig:fplus}
\end{figure}

The gold plated modes $D\to K\ell\nu$ and $D\to\pi\ell\nu$ are most
useful for testing lattice QCD and determining the CKM matrix elements
$|V_{cs}|$ and $|V_{cd}|$. The measurements of the $D\to K$ form
factor are summarized in
Table~\ref{tab:fK}. These numbers agree with the most recent lattice
QCD prediction, $f^K_+(q^2=0)=0.747\pm 0.019$~\cite{Na:2010uf}. Theoretical
calculations based on LQCD also reproduce the form factor shape at
finite values of $q^2$, as shown in Fig.~\ref{fig:fplus}. As mentioned
above, the form factor normalization from lattice QCD can be used to
determine the CKM matrix elements. Using the measurement from
CLEO-c~\cite{Besson:2009uv} and the form factors from
Refs.~\cite{Na:2010uf, fnallqcd}, we obtain $|V_{cd}| = 0.234\pm
0.007\pm 0.002\pm 0.025$ and $|V_{cs}| = 0.963\pm 0.009\pm 0.006\pm
0.024$, where the third uncertainties are from the LQCD calculation of
$f_+(0)$. On $|V_{cs}|$, the lattice error is thus 3\% compared to an
experimental uncertainty of 1\%. For $|V_{cd}|$, the lattice error is
about 10\% while the experimental error amounts to 3\%.
\begin{table}[h]
\caption{Measurements of the $D\to K$ form factor assuming CKM
  unitarity. The uncertainties are statistical and systematic,
  respectively. Third error quoted by BaBar corresponds to the
  uncertainty in external inputs.} \label{tab:fK}
\begin{center}
\begin{tabular}{ccccc}
\hline\hline
Experiment & $f^K_+(q^2=0)$ \\
\hline
Belle~\cite{Widhalm:2006wz} & $0.695\pm 0.007\pm 0.022$\\
BaBar~\cite{Aubert:2007wg} & $0.727\pm 0.007\pm 0.005\pm 0.007$\\
CLEO-c~\cite{Besson:2009uv} & $0.739\pm 0.007\pm 0.005$\\
\hline\hline
\end{tabular}
\end{center}
\end{table}

In addition, CLEO-c has studied the $D\to V\ell\nu$~modes $D\to\rho
e\nu$ and $D^+\to\eta/\eta'/\phi e^+\nu$~\cite{Yelton:2010js}. CLEO-c
has also measured exclusive semileptonic decays of
$D_s$~\cite{:2009cm} and the inclusive semileptonic rates of $D^0$, $D^+$ and $D_s^+$ ~\cite{Asner:2009pu}. Two BaBar analyses study the decays
$D_s^+\to K^+K^-e^+\nu$ and $D^+\to
K^-\pi^+e^+\nu$~\cite{Aubert:2008rs,delAmoSanchez:2010fd}.

\section{\boldmath Leptonic $D$ and $B$ decays}

The leptonic decay constants of $D$ and $B$ mesons can be calculated
using lattice QCD and serve as important inputs to flavor physics
studies.  Again, three groups have recent results for these
quantities: the Fermilab Lattice and MILC Collaborations, HPQCD, and
the ETM Collaboration.  All results are summarized in
Table~\ref{tab:fDB}, along with the world averages as determined by
\cite{Laiho:2009eu}.  In the determination of these lattice averages
correlations are taken into account, the results obtained with a
quenched strange quark are not included, and the PDG prescription for
inflating errors for discrepant results is applied.

\begin{table}[h]
\caption{Lattice results (in MeV) for heavy-light decay constants.}
  \label{tab:fDB}
\begin{center}
\begin{tabular}{ccccc}
\hline\hline
  Analysis       & $f_D$ &$f_{D_s}$ & $f_B$ & $f_{B_s}$ \\ \hline
HPQCD \cite{Davies:2010ip, Gamiz:2009ku}  & $213 \pm 4$ & $248.0 \pm 2.4$ & $190 \pm 13$ & $231\pm15$ \\
FNAL/MILC \cite{Simone:2010zz} & $220\pm9$ & $261\pm 9$ & $212\pm8$ & $256\pm8$ \\
ETMC(quenched strange) \cite{Blossier:2009bx, Blossier:2009gd} & $197\pm 9$ & $244\pm 8$ & $191\pm14$ & $243\pm14$ \\
Average \cite{Laiho:2009eu} & $213.9\pm4.2$ & $248.9\pm3.9$ & $205\pm12$ & $250\pm12$ \\
\hline\hline
\end{tabular}
\end{center}
\end{table}

In the charm sector, the CKM matrix elements $|V_{cs}|$ and $|V_{cd}|$
are strongly constrained by CKM unitarity and leptonic $D$~decays thus
allow to probe predictions of the charm decay constants $f_D$ and
$f_{D_s}$ from lattice QCD assuming the Standard
Model. CLEO-c~\cite{:2008sq,Naik:2009tk} uses its
$E_\mathrm{cm}=3.770$~GeV and $E_\mathrm{cm}=4.170$~GeV data samples
to study $D$ and $D_s$~leptonic decays, respectively. Again, the
analysis strategy is to reconstruct a hadronic final state of the
second charmed meson in the event. Belle~\cite{:2007ws} measures
$D^+_s\to\mu^+\nu$ using a 548~fb$^{-1}$ data sample. In this
analysis, $D_s$~mesons are inclusively reconstructed in events of the
type $e^+e^-\to D^*_s D^{\pm,0}K^{\pm,0}X$ and the $D_s$ 4-momentum is
determined from the recoil system. A recent BaBar
analysis~\cite{delAmoSanchez:2010jg} uses a similar
technique to measure $D^+_s\to\mu^+\nu$ and $D^+_s\to\tau^+\nu$. The
numerical results of these analyses are summarized in
Table~\ref{tab:fDexp}. Comparing these numbers to lattice QCD theory
(HPQCD~\cite{Davies:2010ip}, FNAL/MILC~\cite{Simone:2010zz}), one
finds good agreement for $f_D$. For $f_{D_{s}}$, there is a slight
$2\sigma$ tension between experiment and the new HPQCD prediction.
\begin{table}[h]
\caption{Measurements of the leptonic $D$~decay constants. The
  uncertainties are statistical and systematic,
  respectively.} \label{tab:fDexp}
\begin{center}
\begin{tabular}{ccccc}
\hline\hline
Experiment & $f_D$ (MeV) & $f_{D_s}$ (MeV)\\
\hline
CLEO~\cite{:2008sq,Naik:2009tk} & $206.7\pm 8.5\pm 2.5$ & $259.0\pm
6.2\pm 3.0$\\
Belle~\cite{:2007ws} & & $275\pm 16\pm 12$\\
BaBar~\cite{delAmoSanchez:2010jg} & & $258.6\pm 6.4\pm 7.5$\\
\hline\hline
\end{tabular}
\end{center}
\end{table}

Leptonic $B$~decays have been studied at the $B$~factories Belle and
BaBar. The decay $B^+\to\tau^+\nu$ has the largest branching fraction
and is now well established. Increasingly stringent limits are being set on
leptonic decays involving a light lepton (electron or muon),
$B^+\to\ell^+\nu(\gamma)$. The missing neutrino(s) in the final state
require a tagging technique. The analyses either use a hadronic tag,
in which case the hadronic decays of the other $B$~meson in the event are fully
reconstructed, or a semileptonic tag, in which case a charmed meson
$D^{(*)}$ and a high momentum lepton from the other $B$ are required
in the analysis. In summer 2010, Belle has presented a new measurement
of the $B^+\to\tau^+\nu$ branching fraction using the semileptonic tag
technique, $(1.54^{+0.38}_{-0.37}{}^{+0.29}_{-0.31})\times
10^{-4}$~\cite{Hara:2010dk}. BaBar quotes a new measurement using
hadronic tags, $(1.80^{+0.57}_{-0.54}\pm 0.26)\times
10^{-4}$~\cite{:2010rt}. All measurements of
$B^+\to\tau^+\nu$ are compatible and the Heavy Flavour Averaging Group
quotes a combined branching ratio of $(1.64\pm 0.34)\times
10^{-4}$. This value is in agreement with the Standard Model prediction
of $(1.20\pm 0.25)\times 10^{-4}$, calculated using the HPQCD value
for $f_B$ of $190\pm 13$~MeV and the HFAG value for $|V_{ub}|$ of
$(4.32\pm 0.16\pm 0.29)\times 10^{-3}$~\cite{Barlow:2011fu}. However,
if this measurement is included in an overall fit to the CKM unitarity
triangle, a tension appears as these fits prefer lower values of the
$B^+\to\tau^+\nu$~branching ratio.

\section{\boldmath Semi-leptonic $B$ decays and determination of
 $|V_{cb}|$, $|V_{ub}|$}

The most accurate determinations of the matrix elements  $|V_{cb}|$
and  $|V_{ub}|$ are carried out through analysis of semi-leptonic
$b\to u$ and $b\to c$ quark transitions.   As the theoretical and
experimental methods and uncertainties differ depending on whether 
the decay mode is exclusive or inclusive, these offer complementary
ways of determining these matrix elements.  On the theoretical
side, the main challenge in exclusive decays such as $B\to D \ell \nu$ and 
$B \to \pi \ell \nu$ is to determine the non-perturbative form factors
entering expressions for the decay rate, either in lattice QCD or 
light-cone sum rules.  For inclusive decays, the $B\to X_c \ell \nu$
decay width can be reliably calculated using a straight-forward
operator product expansion (OPE), while the experimental cuts needed
in measurements of $B\to X_u \ell \nu$ introduce sensitivity to
non-perturbative shape-functions and the theoretical treatment is more
involved.

\subsection{\boldmath $b$- and $c$-quark masses}

An important input to the determination of the CKM matrix elements $|V_{cb}|$
and $|V_{ub}|$ from semi-leptonic decays are the $b$- and $c$-quark
masses. These heavy-quark masses can either be treated as external input and
taken from lattice or QCD sum rule methods, or determined along with the CKM
matrix elements in global fits for the inclusive decays.  As the pole masses
suffer from renormalon ambiguities of the order of $\delta m_{c,b}\sim
\Lambda_{\rm QCD}$, it is necessary that these determinations be carried out
in short-distance schemes which are free of such ambiguities.  Such schemes
can be divided into two categories: the $\overline{\rm MS}$ scheme, typically
used in QCD sum rule and lattice determinations, and threshold schemes such as
the kinetic, 1S, or shape-function schemes, typically used in global fits of
semi-leptonic $B$ decays into $c$ or $u$ quarks.

In the working-group II session, new results for the heavy-quark masses using
global fits of inclusive semi-leptonic $b$ and $c$ decays were presented, and
will be discussed below in that context.  In addition, there was a dedicated
talk by A.\ Hoang on determining $m_c$ in the $\overline{\rm MS}$ scheme from
QCD sum rules and experimental data from charm production $e^+e^-$ collisions,
based on work performed in \cite{Dehnadi:2011gc}. While the determination of
$m_c$ from QCD sum rules is already in an advanced state, and recent
calculations \cite{Boughezal:2006px, Chetyrkin:2009fv} lead to a value of
$m_c$ with very small errors, the purpose of the study was to re-examine
several aspects of the current analyses.  In particular, the full set of
experimental data was included, and special attention was paid to the
perturbative  error analysis, accounting for all sources of 
scale variation and different ways of expanding the series.  The final
results of the analysis was an $\overline{\rm MS}$ mass of  
\cite{Dehnadi:2011gc}
\begin{eqnarray}
\label{mcfinalalphaswa}
\overline m_c(\overline m_c) & = & 1.277 
\pm (0.006)_{\rm stat}
\pm (0.013)_{\rm syst}
\pm (0.019)_{\rm pert}
\pm (0.009)_{\alpha_s}
\pm (0.002)_{\langle GG\rangle}
\nonumber
\\
& = &
1.277 \, \pm \, 0.026\,\,\mbox{MeV}~,
\end{eqnarray}
where in the first line the first and second errors come from experimental
uncorrelated and correlated uncertainties, respectively, the third error is
the perturbative uncertainty, the fourth reflects the uncertainty in
$\alpha_s(m_Z)$, and the last corresponds to non-perturbative effects from the
gluon condensate.  The central value does not differ significantly from that
found in the analysis of \cite{Chetyrkin:2009fv}, since the differences in 
the experiment and theoretical analyses largely cancel one another. 
However, the perturbative error estimate associated with the truncation
of the series is an order of magnitude larger, leading to a total 
error which is larger by about a factor of two.

\subsection{\boldmath Semi-leptonic $b\to c$ decays and $|V_{cb}|$}

\noindent {\it Exclusive decays}:  The determination of $|V_{cb}|$
from exclusive $B$~decay measurements requires the calculation of
nonperturbative form factors, usually provided by lattice QCD.  This
conference saw an update of the Fermilab/MILC Collaborations' lattice
determination of the $B\to D^*\ell\nu$ form factor at zero recoil
\cite{Bernard:2008dn}.  The improvements in the update are due mainly
to increased statistics and the use of finer lattice spacings.  Their
new determination of the form factor is $F(1)=0.908\pm 0.17$
\cite{Bernard:2008dn}, and taking the latest Heavy Flavor Averaging
Group (HFAG) update of $|V_{cb}|F(1)\times 10^3=36.04\pm0.52$ from
experiment \cite{HFAG:2009}, the new value of $|V_{cb}|$ from
exclusive $B\to D^*\ell\nu$ is $|V_{cb}|=39.7(7)(7)\times 10^{-3}$
\cite{Mackenzie:2010}, where the errors are experimental and
theoretical. The $F(1)$~form factor has also recently been calculated
using zero recoil sum rules, yielding to $F(1)=0.86\pm 0.04$ and thus a larger
value of $|V_{cb}|$ exclusive~\cite{Gambino:2010bp}. Improvements in
the experimental error from semi-leptonic decays may eventually come
from LHCb, as presented at this conference~\cite{Urquijo:2011en}.

\noindent {\it Inclusive decays}: The theoretical tool for
understanding inclusive $B$~decays is the Operator Product Expansion
(OPE) which allows to express the transition amplitude as a double
expansion in $\alpha_s$ and $\Lambda_\mathrm{QCD}/m_b$. In the OPE,
non-perturbative physics is expressed in terms of matrix elements of
local operators, while the Wilson coefficients are
perturbative. Expansions for inclusive observables in $B$~decays are
available in two implementations, the
kinetic~\cite{Benson:2003kp,Gambino:2004qm,Benson:2004sg} and the
1S~scheme~\cite{Bauer:2004ve}. They both include terms up to
$\mathcal{O}(\alpha^2_s\beta_0)$ and $\mathcal{O}(1/m^3_b)$. A near
term improvement is the implementation of the complete two-loop
perturbative corrections in the kinetic scheme~\cite{Gambino:2011fz}.

To obtain $|V_{cb}|$ with a precision of a few percent, the
non-perturbative matrix elements are obtained from a global fit to
experimental moments of inclusive $B$~observables. Currently the
moments of the lepton energy and the hadronic mass in $B\to
X_c\ell\nu$ and the moments of the photon energy spectrum in $B\to
X_s\gamma$ are used. Recent measurements of these quantities were
performed by Belle~\cite{Schwanda:2008kw} and
BaBar~\cite{Aubert:2009qda}. These analyses measure the inclusive
spectra in hadronically tagged events and employ various techniques to
correct for the distortions due to the measurement device.

The global fit to the experimental data is now performed by HFAG to
combine data from different experiments and obtain optimal
determinations of $|V_{cb}|$ and the $b$-quark mass $m_b$ from
inclusive $b\to c$~decays. This fit uses a total of 66
measurements -- 29 from BaBar, 25 from Belle and 12 from other
experiments. The results in the kinetic and the 1S schemes are given
in Tables~\ref{tab:kin} and \ref{tab:1s}. In both cases, the results
with all moments and with $B\to X_c\ell\nu$~moments only are quoted.
There is a $\sim2\sigma$ tension between the inclusive and exclusive determinations of $|V_{cb}|$.
\begin{table}
  \caption{Results of the HFAG global fit in the kinetic scheme. The
    errors quoted are the results of the fit, where the covariance
    matrix includes experimental and estimated theoretical
    uncertainties. On $|V_{cb}|$, there are additional uncertainties
    from the $B$~lifetime and from an additional theoretical
    uncertainty of 1.4\% in the expression of the semileptonic width,
    respectively.}
  \label{tab:kin}
  \begin{center} \begin{tabular}{c|cccc}
    \hline \hline
    Input & $|V_{cb}|$ (10$^{-3}$) & $m^\mathrm{kin}_b$ (GeV) &
    $\chi^2/$ndf.\\
    \hline
    all moments & $41.85\pm 0.42\pm 0.09\pm 0.59$ & $4.591\pm 0.031$ &
    29.7/59\\
    $X_c\ell\nu$ only & $41.68\pm 0.44\pm 0.09\pm 0.58$ & $4.646\pm
    0.047$ & 24.2/48\\
    \hline \hline
  \end{tabular} \end{center}
\end{table}
\begin{table}
  \caption{Results of the HFAG global fit in the 1S scheme. The
    errors quoted are the results of the fit, where the covariance
    matrix includes experimental and estimated theoretical
    uncertainties.}
 \label{tab:1s}
  \begin{center} \begin{tabular}{c|cccc}
    \hline \hline
    Input & $|V_{cb}|$ (10$^{-3}$) & $m^{1S}_b$ (GeV) &
    $\chi^2/$ndf.\\
    \hline
    all moments & $41.87\pm 0.25$ & $4.685\pm 0.029$ & 32.0/57\\
    $X_c\ell\nu$ only & $42.31\pm 0.36$ & $4.619\pm 0.047$ & 24.2/46\\
    \hline \hline
  \end{tabular} \end{center}
\end{table}

\subsection{\boldmath Semi-leptonic $b\to u$ decays and $|V_{ub}|$}

\noindent {\it Exclusive decays}:  
The $B\to \pi \ell \nu$ decay rate is proportional to 
the combination $|V_{ub}|^2 f^2_+(q^2)$, where $f_+(q^2)$ is 
a nonperturbative form factor.   To determine $|V_{ub}|$ 
from exclusive semi-leptonic decays thus requires measurements
of the decay rates, and calculations of the form factor.

In the working-group session an update on the determination of the form factor
using light-cone sum rules (LCSR) was given by P.~Ball \cite{Ball}. The LCSR
calculations for the form factor are already in a mature state, and the focus
was on a new calculation of the ${\cal O}(\alpha_s^2 \beta_0)$ terms used in
the perturbative part of the sum rule.  The numerical effect of such
corrections turns out to be small, which can be taken as an indication that the
current results in LCSR \cite{Ball:2004ye, Duplancic:2008ix} are stable under
higher-order radiative corrections. After the workshop, a new LCSR
result appeared~\cite{Khodjamirian:2011ub}.

As sum rule calculations are subject to theoretical uncertainties
which are difficult to quantify, much effort has been put into 
calculating these quantities  using the model-independent methods
of lattice QCD.  While new results for the form factor $f_+(q^2)$
used in exclusive $b\to u$ transitions were not presented in 
this workshop, the existing results from Refs.~\cite{Dalgic:2006dt} and
\cite{Bernard:2009ke} are used in experimental analyses.  

New measurements from BaBar~\cite{delAmoSanchez:2010zd,:2010uj} and
Belle~\cite{Ha:2010rf} were presented at this workshop, along with
updates of $|V_{ub}|$ determinations based on calculations of the form
factors mentioned above. All analyses use an untagged technique,
\textit{i.e.}, no requirements are made on the second $B$~meson in the event.
Different methods are used to determine the values of $q^2$,
however. The BaBar and Belle results for $|V_{ub}|$ are shown in
Tables~\ref{vubtable} and \ref{tab:vub}, respectively.
\begin{table}[h]
\caption[]{\label{vubtable} Values of $|V_{ub}|$ ($10^{-3}$) derived
  from different $B\to\pi\ell\nu$ form factor calculations and two
  recent BaBar analyses~\cite{delAmoSanchez:2010zd,:2010uj}.}
\begin{center}
%\begin{normalsize}
\begin{tabular}{ccccc}
\hline\hline
  & $q^2$ (GeV$^2$) & Ref.~\cite{delAmoSanchez:2010zd} &
  Ref.~\cite{:2010uj} & Average\\
\hline
HPQCD & $>16$ & $3.28\pm 0.13\pm 0.15{}^{+0.57}_{-0.37}$ &
$3.21\pm 0.17{}^{+0.55}_{-0.36}$ & $3.23\pm 0.09\pm
0.13{}^{+0.57}_{-0.37}$\\
FNAL & $>16$ & $3.14\pm 0.12\pm 0.14{}^{+0.35}_{-0.29}$ &
$2.95 \pm 0.31$ & $3.09\pm 0.08\pm 0.12{}^{+0.35}_{-0.29}$\\
LCSR & $<12$ & $3.70\pm 0.07\pm 0.08{}^{+0.54}_{-0.39}$ &
$3.78\pm 0.13{}^{+0.55}_{-0.40}$ & $3.72\pm 0.05\pm 0.09{}^{+0.54}_{-0.39}$\\
\hline\hline
\end{tabular}
%\end{normalsize}
\end{center}
\end{table}
\begin{table}[h]
\caption{Values for $|V_{ub}|$ extracted from the Belle
  data~\cite{Ha:2010rf} using different predictions of the partial
  width $\Delta\zeta$.}
\label{tab:vub}
\begin{center}
\begin{tabular}{lccc}
%{@{\hspace{0.1cm}}l@{\hspace{0.1cm}}@{\hspace{0.1cm}}c@{\hspace{0.4cm}}c@{\hspace{0.4cm}}c@{\hspace{0.05cm}}}
\hline \hline
& $q^2$ (GeV$^2$) & $\Delta \zeta$ (ps$^{-1}$) & $|V_{ub}|$ (10$^{-3}$)\\
\hline
HPQCD   &  $>$ 16         & 2.07 $\pm$ 0.57            & 3.55 $\pm$ 0.09 $\pm$ 0.09 $^{+0.62}_{-0.41}$ \\
FNAL/MILC   &  $>$ 16         & 1.83 $\pm$ 0.50            & 3.78 $\pm$ 0.10 $\pm$ 0.10 $^{+0.65}_{-0.43}$ \\
LCSR      &  $<$ 16         & 5.44 $\pm$ 1.43            & 3.64 $\pm$ 0.06 $\pm$ 0.09 $^{+0.60}_{-0.40}$ \\
\hline \hline
\end{tabular}
\end{center}
\end{table}
Results from a model-independent determination of $|V_{ub}|$ obtained by
simultaneously fitting the branching fraction data and the MILC lattice QCD
form-factor after transforming to the so-called
``z-parameterization''~\cite{Bailey:2008wp} were also presented. The result of
the fit to the BaBar data reads $|V_{ub}|= (2.95 \pm 0.31) \times
10^{-3}$, while the Belle data gives $|V_{ub}| = ( 3.43 \pm 0.33)
\times 10^{-3}$.

\vspace{0.5cm}

\noindent {\it Inclusive decays}: In principle, the same methods used for
inclusive semi-leptonic decays into charm quarks described above can be used to
determine $|V_{ub}|$ from $B\to X_u \ell \nu$ decays.  However, in this case
experimental cuts are required to suppress the background from decays into
charm, and for very restrictive cuts the local OPE does not apply and theory
predictions are sensitive to non-perturbative shape functions. Moreover, the
dependence of the partial branching fractions in such a region of phase space
depends much more strongly on $m_b$ than in the total inclusive rate, so
parametric uncertainties become large.

Current determinations by HFAG \cite{Asner:2010qj} of $|V_{ub}|$ from
inclusive decays are based on theory frameworks referred to as BLNP
\cite{Bosch:2004th}, DGE \cite{Andersen:2005mj}, GGOU \cite{Gambino:2007rp},
and ADFR \cite{Aglietti:2007ik}.   Basically, these methods use a
different set of theoretical assumptions, but BLNP, DGE, and GGOU are linked
by the fact that they in one way or another produce the local OPE result for
the total rate, which is however not the case for ADFR.  
More details can be found, for instance, in the report of the 
previous CKM workshop \cite{Antonelli:2009ws}.

Compared to the previous CKM workshop, theoretical progress within the
BLNP framework was made in \cite{Greub:2009sv}, which included the
next-to-next-to-leading order (NNLO) perturbative corrections to the
leading term in the $1/m_b$ expansion within that framework.  These
corrections stabilize the dependence on the perturbative matching
scales, and tend to raise the value of $|V_{ub}|$ compared to the
current implementation of BLNP used by HFAG, which is based on
next-to-leading-order calculations.  Since these corrections are part
of the OPE prediction for the triple differential decay spectrum,
namely the virtual plus real emission contributions in the soft and
collinear limits, they could also be included in the GGOU and DGE
frameworks.  Whether this would be beneficial depends on whether the
NNLO contributions from hard real gluon emission, which are power
suppressed and not included in \cite{Greub:2009sv}, give significant
contributions in the regions of phase space where experimental
measurements are made. Further efforts in understanding the structure
of such power corrections  within the BLNP framework were made
through the calculation of the subleading jet functions at ${\cal
  O}(\alpha_s)$ in \cite{Paz:2009ut}. This set of perturbative power
corrections appears in a convolution with the leading-order shape
function and is suppressed by $\alpha_s/m_b$ compared to the leading
term.  The implementation of the subleading jet functions in a
numerical analysis within the BLNP framework should be relatively
straightforward, but was not yet performed.

Progress has also been made in estimating contributions from weak
annihilation.  Although at the level of the total rate weak annihilation can
be treated within the OPE and is of the order $1/m_b^3$, its
calculation at the level of differential decay spectrum is model
dependent and its effect on partial decay rates used in the extraction
of $|V_{ub}|$ is more uncertain. In Refs.~\cite{Ligeti:2010vd,
  Gambino:2010jz}, recent CLEO data on semi-leptonic $D$ and $D_s$
decay \cite{Asner:2009pu} and heavy-quark symmetry
\cite{Voloshin:2001xi} were used to estimate the potential effect of weak
annihilation on extractions of $|V_{ub}|$.  The main result of both of these
studies is that the weak annihilation contribution on the fully inclusive
measurement is only at most a  2\% effect at the level of the total rate.
 
A further development has been the advent of the SIMBA collaboration, which
aims at extracting $|V_{ub}|$ within the context of a global fit to $B\to
X_s\gamma$ and $B\to X_u \ell \nu$ decays. A talk on the status of the SIMBA
programme was given by F.~Tackmann and is summarized in
\cite{Bernlochner:2011di}.  Roughly speaking, the theoretical framework
underlying the approach is similar to that used in BLNP, although it differs
in the treatment of non-perturbative shape functions \cite{Ligeti:2008ac}.
The BLNP and GGOU approaches take a model for the shape function which is
motivated by the shape of the $B\to X_s\gamma$ photon energy spectrum, and
scan over many possibilities to determine uncertainties associated with this
object.  The intention of the SIMBA collaboration, on the other hand, is to
use the available data from $B\to X_s\gamma$ and $B\to X_u \ell \nu$ decays
to constrain the shape function, $|V_{ub}|$, and $m_b$ in a global fit.
Preliminary results were given for a fit of the shape-function, $m_b$, and
$|C_7^{\rm incl} V_{tb}V^*_{ts}|$ (the normalization factor multiplying the
photon energy spectrum) from the photon energy spectrum in $B\to X_s\gamma$
decays; a full study including also $B\to X_u\ell \nu$ decays is in progress.
Since the goal is to extract not only the normalization of the decay rate,
proportional to $|V_{ub}|$, but also its shape, determined at leading
power by a single non-perturbative shape function, it would be very useful for
this effort if the data on $B\to X_u\ell \nu$ decays were given in the form of
differential spectra, rather than just partial decay rates for a few different
choices of kinematical cuts.

On the experimental side, an update was given by C.~Bozzi \cite{Bozzi:2011aj}
on the status of measurements by Belle and Babar, and also the preliminary
updates of inclusive $|V_{ub}|$ as extracted by HFAG.  In addition to analyses
which perform measurements in the endpoint region of the lepton energy
spectrum, both collaborations also have measurements of partial rates with
cuts on variables such as the hadronic invariant mass, the $q^2$ of the lepton
pair, or the light-cone momentum $P_+ = E_X - |\vec{p}_X|$ of the hadronic
system, which are carried out using recoil techniques \cite{Aubert:2007rb,
:2009tp}.  The preliminary results for the value of $|V_{ub}|$ extracted for a
number of such measurements within the different theoretical frameworks can be
found in \cite{Bozzi:2011aj}; the average over all measurements is shown in
Table~\ref{tab:VUB}.  As already mentioned, the theory predictions for the
partial rates depend strongly on the heavy-quark parameters: the numbers in
the table correspond to those determined by a global fit in the kinetic
scheme, translated to the scheme needed by each method, where both
$b\rightarrow c \ell \nu$ and $b \rightarrow s \gamma$ moments are used,
giving $m_b(\mathrm{kin})=4.591\pm 0.031$ GeV,
$\mu_{\pi}^2(\mathrm{kin})=0.454\pm 0.038$ GeV$^2$.  There is an
approximately $3\sigma$ discrepancy between the exclusive and
inclusive determinations of $|V_{ub}|$.  Further work is needed to
understand this discrepancy.

\begin{table}[tbp]
\caption{Results for $|V_{ub}|\times 10^3$ obtained with four theoretical
 calculations, taken from \cite{Bozzi:2011aj}. The uncertainties are
 experimental ({\em{i.e.}}  sum of statistical and experimental systematic)
 and theoretical, respectively.} \label{tab:VUB}
\begin{center}
\vspace{0.1in}
\footnotesize
\begin{tabular}{llcccc} 
\hline \hline
 	&  & BLNP 		& DGE 		& GGOU 		& ADFR
\\
\hline
\multicolumn{2}{c}{Average} 	& $4.30\pm 0.16 ^{+0.21}_{-0.23}$ & $4.37\pm 0.15 ^{+0.17}_{-0.16}$	
				& $4.30\pm 0.16 ^{+0.13}_{-0.20}$ & $4.05\pm
 	0.13 ^{+0.24}_{-0.21}$ \\
\hline \hline
\end{tabular}
\end{center}
\end{table}

%\begin{itemize}
%\item Comment on discrepancy with exclusive determinations
%\item Not sure where to put Urquijo's talk
%\end{itemize}

%%%%%%%%%%%%%%%%%%%%%%%%%%%%%%%%%%%%%%%%%%%%%%%%%%%%%%%%%%%%%%%%%%%%%%%%%
%%
%%   use this format to include an .eps figure into your paper
%%
% \begin{figure}[htb]
% \centering
% \includegraphics[height=1.5in]{magnet}
% \caption{Plan of the magnet used in the mesmeric studies.}
% \label{fig:magnet}
% \end{figure}
%%%%%%%%%%%%%%%%%%%%%%%%%%%%%%%%%%%%%%%%%%%%%%%%%%%%%%%%%%%%%%%%%%%%%%%%%%%

%%%%%%%%%%%%%%%%%%%%%%%%%%%%%%%%%%%%%%%%%%%%%%%%%%%%%%%%%%%%%%%%%%%%%%%%%
%%
%%   use this format to include a LaTeX table  into your paper
%%
% \begin{table}[t]
% \begin{center}
% \begin{tabular}{l|ccc}  
% Patient &  Initial level($\mu$g/cc) &  w. Magnet &  
% w. Magnet and Sound \\ \hline
%  Guglielmo B.  &   0.12     &     0.10      &     0.001  \\
%  Ferrando di N. &  0.15     &     0.11      &  $< 0.0005$ \\ \hline
% \end{tabular}
% \caption{Blood cyanide levels for the two patients.}
% \label{tab:blood}
% \end{center}
% \end{table}
%%%%%%%%%%%%%%%%%%%%%%%%%%%%%%%%%%%%%%%%%%%%%%%%%%%%%%%%%%%%%%%%%%%%%%%%%%%

\Acknowledgements
We thank the organizers for a well-organized conference and the
participants of Working Group II for helpful discussions about their
work.

\end{document}